# Using Analogy to Solve a Three-Step Physics Problem

Shih-Yin Lin and Chandralekha Singh

*Department of Physics and Astronomy, University of Pittsburgh, Pittsburgh, PA 15260, USA*

**Abstract.** In a companion paper, we discuss students' ability to take advantage of what they learn from a solved problem and transfer their learning to solve a quiz problem that has different surface features but the same underlying physics principles. Here, we discuss students' ability to perform analogical reasoning between another pair of problems. Both the problems can be solved using the same physics principles. However, the solved problem provided was a two-step problem (which can be solved by decomposing it into two sub-problems) while the quiz problem was a three-step problem. We find that it is challenging for students to extend what they learned from a two-step problem to solve a three-step problem.

## INTRODUCTION

In a companion paper[1], we discussed how introductory physics students performed when they were explicitly asked in the recitation quiz to learn from the solution to a problem (the solved problem) and transfer their learning to solve another problem (the quiz problem) that had the same underlying physics but different surface features. In that study, both the quiz problem and the solved problem are two-step problems that can be solved by applying two physics principles in the same order. Here, we present a similar study to examine students' ability to perform analogical reasoning between another pair of problems. Both the quiz problem and the solved problem can again be solved using the same principles. However, unlike the companion study, the quiz problem in this study can be solved by decomposing it into three sub-problems (each of which involves a single physics principle) while the solved problem can be solved by decomposing it into two sub-problems. The goal is to not only examine students' ability to discern the similarities between the two problems and identify the relevant principles involved, but also to explore if students are able to extend what they learned from a problem having a two-step solution to solve a problem having a three-step solution. In order to solve the three-step quiz problem, students must learn to systematically decompose a multi-step problem into several sub-problems, and more importantly, to realize how the consecutive sub-problems are connected.

## METHODOLOGY

Introductory physics students from a calculus-based and an algebra-based course (398 total) participated in this study. Recitation classrooms in both courses were distributed into one comparison group and three intervention groups (intervention groups 1, 2 and 3). Students in all groups were asked to solve the same quiz problem in the recitation. Students in the different intervention groups were provided with a solved problem in addition to the quiz problem; they were explicitly asked to point out the similarities between the two problems and explain whether they can use the similarities to solve the quiz problem. The solved problem provided was about a person who took a running start on the level section of a track, jumped onto a stationary snowboard and then went up with the snowboard. The problem asked for the minimum speed at which the person should run in order to go up to at least a certain height given the masses of the person and the snowboard, assuming the friction can be ignored. A detailed solution explaining why and how each principle is applicable was attached.

The quiz problem to be solved, on the other hand, was about two small putty spheres of equal mass hanging from the ceiling on massless strings of equal length. Putty A was raised to a height $h_0$ and released. After putty A collided with putty B, which was initially at rest, the two putties stuck and swung together to a maximum height $h_f$. Students were asked to find the maximum height $h_f$ in terms of $h_0$. This quiz problem can be divided into 3 steps (putty A going down, the collision process, and the two putties going up together) with the last two steps, which can be solved using the principles of conservation of momentum (CM) and conservation of total mechanical energy (CME) respectively, being directly analogous to the solved problem. The 1st step in which putty A goes down is similar to the 3rd step in which two putties go up and can be solved using the principle of CME. The same problems have also been used in other research [2, 3]. Although the quiz problem may seem easy to a physics expert, it is quite challenging for the students because solving the problem correctly requires the ability to decompose the problem into several suitable sub-problems that can be tackled one at a time. What's more, students need to have a clear picture of how a variable in one sub-problem is related to a variable in a consecutive sub-problem. The existence of the additional step in the quiz problem as compared to the solved problem therefore makes it more challenging to make an analogy and transfer the learning from one problem to another.

Three different interventions were implemented to help students learn via analogical reasoning. In intervention 1, students were explicitly told in the recitation quiz to take the first 10 minutes to learn from the solution of the snowboard problem. Then, they turned in the solved problem and were given two problems to solve: one was the same as the problem they just browsed over (the snowboard problem), and the other was the putty problem. In intervention 2, students were asked to solve the putty problem on their own without being provided the solution to the snowboard problem. After 10 minutes, they turned in their solutions, and then were provided the solved snowboard problem to learn from. Then, they were asked to redo the quiz problem. We hypothesize that by reproducing the solved example or struggling with the quiz problem first, students may learn from the solved example better. A more complete description of the objectives of interventions 1 and 2 can be found in the companion paper [1].

Intervention 3 was designed with explicit guidance to direct students' attention to the principles involved in both problems. Students in this group were given both the quiz problem and solved problem together. They were explicitly told that similar to the snowboard problem, the putty problem could be solved using the principles of CME and CM and they might have to use CME twice to find the height $h_f$ in terms of $h_0$. Our previous research indicates that if students are simply asked to learn from the solved example of the two-step snowboard problem to solve the three-step putty problem without any additional scaffolding, they have great difficulty dealing with the additional step in the putty problem. It was unclear whether the students realized that the putty problem can be decomposed into three sub-problems. We expected that by providing a hint about using the CME twice, they may be able to approach the problem more systematically.

Students' performance on the quiz was later graded using a rubric. An inter-rater reliability of at least 80% was achieved when two researchers scored independently a sample of 10% of the students. The rubric, which has a full score of 10 points, can be divided into 2 parts based upon the two principles involved. Table 1 shows the summary of the rubric of the putty problem, with 4 and 6 points devoted to the principles of CM and CME, respectively. The rubric for the snowboard problem is slightly different but similar. Students' performance in the interventions 1, 2 and 3 were analyzed and compared with the performance of the comparison group which solved the putty problem without any scaffolding provided. In addition to the whole group average, we also examine how students with a particular expertise perform by further classifying the students in a course as top, middle, bottom and none based on their scores on the final exam ("none" means they didn't take the final exam). By doing so, we can gain a better understanding of the impact of different interventions on students with different expertise.

**TABLE 1.** Summary of the rubric for the putty problem.

| Description | Scores |
|---|---|
| Conservation of Mechanical Energy in the 1st and 3rd sub-problems (6 points) | Invoking physics principle: 2 points (1 point for each sub-problem) |
| | Applying physics principle: 4 point (2 points for each sub-problem) |
| Conservation of Momentum in the 2nd sub-problem (4 points) | Invoking: 1 point |
| | Applying: 1 point |
| | Relevant to the final answer: 2 points |

# RESULTS

Tables 2 and 3 present students' average scores on the putty problem in the calculus-based and algebra-based courses. The average scores of the comparison group students in the two courses were 6.4/10 and 2.3/10 respectively, indicating that the putty problem is challenging for the calculus-based students and almost impossible to the algebra-based student. Other research [3] has shown that students have difficulty in identifying all the relevant principles for solving the putty problem, which is consistent with our finding here. We found that forgetting to invoke the principle of CM is students' most common mistake on the putty problem when no scaffolding was provided. Many of them simply relate the initial potential energy of putty A (when it is raised to an initial height) to the final potential energy of putty A and B (when both of them reach the maximum height) and come up with an expression $m_A g h_o = (m_A + m_B) g h_f$ without considering the intermediate collision process. Other students took into account the intermediate process but still came up with a similar answer $m_A g h_o = 1/2\ mv^2 = (m_A+m_B)gh_f$. (Here, depending on the student, $m$ and $v$ could stand for the mass and the speed of putty A right before the collision, or the mass and the speed of both putties together right after the collision.) These answers suggest that students have the notion of the mechanical energy being conserved during the whole process. They didn't notice that there was an inelastic collision process involved and only the momentum, not the mechanical energy, is conserved during an inelastic collision. Moreover, students incorrectly combined the various sub-problems into one without systematically approaching different parts. Other common mistakes are listed in Table 4 which show the great difficulty students have about the collision process and the various velocities involved.

**TABLE 2.** Students' Average Scores out of 10 on the putty problem in the calculus-based course. The number of students in each case is shown in parentheses.

|        | compare  | Intv 1   | Intv 2 before | Intv 2 After | Intv 3   |
|--------|----------|----------|---------------|--------------|----------|
| Top    | 8.2 (13) | 9.2 (13) | 6.1           | 8.4 (13)     | 8.2 (19) |
| Middle | 6.8 (12) | 6.1 (10) | 6.9           | 8.4 (10)     | 6.8 (35) |
| Bottom | 3.9 ( 9) | 3.8 (14) | 3.3           | 5.2 (12)     | 5.6 (20) |
| None   | 5.3 ( 4) |          |               |              | 2.5 ( 2) |
| All    | 6.4 (38) | 6.3 (37) | 5.3           | 7.3 (35)     | 6.7 (76) |

**TABLE 3.** Students' Average Scores out of 10 on the putty problem in the algebra-based course. The number of students in each case is shown in parentheses.

|        | compare  | Intv 1   | Intv 2 before | Intv 2 After | Intv 3   |
|--------|----------|----------|---------------|--------------|----------|
| Top    | 3.8 (10) | 5.3 (27) | 4.6           | 7.3 (21)     | 6.2 (15) |
| Middle | 1.9 (19) | 3.3 (11) | 1.9           | 4.2 (17)     | 5.3 (17) |
| Bottom | 1.8 (16) | 4.5 ( 8) | 1.2           | 4.5 (24)     | 4.2 (16) |
| None   | 1.3 ( 3) | 1.5 ( 2) | 10            | 1.5 ( 2)     | 5.5 ( 4) |
| All    | 2.3 (48) | 4.5 (48) | 2.5           | 5.2 (64)     | 5.2 (52) |

Even though students in the intervention groups 1, 2 and 3 received the solved example of the snowboard problem and other scaffoldings to help them solve the putty problem, students' performance in the various interventions (see Tables 2 and 3) doesn't show great improvement. In fact, none of the intervention groups in the calculus-based course show a performance statistically different from that of the comparison group. Comparing intervention 2 students' scores in the calculus-based course before and after the scaffolding shows that students did improve after learning from the solved example; however, the improvement is not large enough to make a statistical difference from the comparison group. Although the algebra-based students in all three intervention groups did perform significantly better (with p value < 0.05) than the comparison group students, there is still much room for improvement.

A comparison with the result in the companion paper indicates that it is much harder for students to extend what they learned from a two-step problem to solve a three-step problem than simply going from a two-step problem to another two-step problem. Even though the problems in the companion paper requires the application of Newton's 2$^{nd}$ Law in the non-equilibrium situation with a centripetal acceleration involved (which is generally considered a more difficult problem), students performed reasonably well in transferring what they learned from the solved problem to solve the quiz problem. We believe that the difference between these two results lies in the fact that decomposing a problem appropriately into several sub-problems and figuring out how the different sub-problems are connected is extremely difficult for the students. With the existence of an additional step in the quiz problem, students can no longer map the solved problem directly to the quiz problem. They have to learn from the solved example and understand the circumstances for which each principle is applicable so that they can systematically decompose the problem into several sub-problems that can be dealt with one at a time with a single principle. More importantly, they have to carefully think through the fact that the final speed they found in the 1$^{st}$ sub-problem when putty A goes down will become the initial speed for the collision process in the 2$^{nd}$ sub-problem. Similarly, the final speed of putties A and B together right after the collision in the 2$^{nd}$ sub-problem will become the new initial speed in the 3$^{rd}$ sub-problem when the two putties swing together to their maximum height. If students don't have a holistic picture of the entire process of how the speeds in different sub-problems are connected and if students don't use appropriate notation for the various speeds involved, they are likely to make mistakes. As Table 4 shows, students struggled with the putty problem and had the mistake of erroneously mixing up several processes into one. Such problems were more commonly found in the algebra-based course than in the calculus-based course. Providing students with the solution to the snowboard problem doesn't necessarily help students in applying these principles correctly.

**TABLE 4.** Summary of students' common mistakes.

| Description of Students' mistake | Example of students' answer |
|---|---|
| Mechanical Energy is conserved during the whole process (and combine 2 sub-problems into 1) | $m_A g h_o = (m_A+m_B)gh_f$ |
|  | $m_A g h_o = ½ (m_A+m_B)v^2$, $½ (m_A+m_B)v^2 = (m_A+m_B)gh_f$ |
| velocity is the same before and after the collision | $m_A g h_o = ½ m_A v^2 \Rightarrow v^2 = 2g\ h_o$, $½ (m_A+m_B)v^2 = (m_A+m_B)gh_f \Rightarrow h_f = v^2/2g = h_o$ |
| Combine 3 sub-problems into 1 | $m_A g h_o + ½ m_A v_A^2 = (m_A+m_B)gh_f + ½ (m_A+m_B)v_f^2$ |
| Combine 3 sub-problems into 1 | $v_{A+B} = m_A v_A/(m_A+m_B)$, $m_A g h_o + ½ (m_A+m_B)v_{A+B}^2 = (m_A+m_B)\ g\ h_f$ |

**TABLE 5.** Average scores out of 10 on the snowboard problem (solved problem) and the putty problem (quiz problem) for intervention 1 in the algebra- and calculus-based courses.

|  | Solved Problem | | Quiz Problem | |
| --- | --- | --- | --- | --- |
|  | Calculus | Algebra | Calculus | Algebra |
| Top | 9.9 | 8.8 | 9.2 | 5.3 |
| Middle | 9.9 | 6.8 | 6.1 | 3.3 |
| Bottom | 8.9 | 9.4 | 3.8 | 4.5 |
| None |  | 5 |  | 1.5 |
| All | 9.5 | 8.3 | 6.3 | 4.5 |

Table 5 shows intervention 1 students' average scores on the snowboard problem immediately after learning from and returning its solution. Students' performance on the quiz problem is also listed for comparison. Although students performed well on the snowboard problem when they were asked to reproduce it, this score turned out to be somewhat superficial when it comes to transfer. An average drop of 3.2 and 3.8 out of 10 on the putty problem were found in the calculus and algebra-based courses, respectively. As discussed earlier, it was not easy for students to transfer what they learned from a two-step solved problem provided to solve the three-step quiz problem. Even students in intervention 3 who received an explicit instruction on "applying the conservation of mechanical energy twice" had great difficulty figuring out the correct process to solve the quiz problem. Examination of students' work in intervention 2 before and after they read the snowboard problem indicates that the solved example is most useful in helping students invoke the correct principles, in particular the principle of CM, which was more likely to be ignored by the students who didn't receive the solved snowboard problem. However, even if students realized that the principle of CM should be applied to the collision process, they didn't necessarily discern the relevance of this principle to the final answer. Many students didn't make use of the CM principle to relate $h_f$ to $h_o$; they still used their original idea (e.g., that the mechanical energy is conserved during the whole process) to solve for the final answer even though they have found that the speed of putties stuck together after the collision is half of the initial speed of putty A right before the collision by applying the principle of CM successfully. Although some improvement is seen among students who are able to take advantage of the snowboard problem and successfully map the last two sub-problems of the putty problem to it, many of them didn't know what to do with the 1st sub-problem that was not included in the solved problem and some of them just left it unattended. Other students who struggled even more weren't able to discern the three-step nature of the quiz problem or the correspondence between the quiz and solved problems; they again mistakenly combined several processes into one after browsing over the solved problem.

## DISCUSSION

Students' major difficulties on the putty problem include the challenges in invoking the CM principle and applying the principle of CME correctly. It is easier to learn from the solved problem that the CM principle should be invoked in the quiz problem than to learn to apply the CME correctly in a three-step problem. In both the algebra-based and calculus-based courses, most students who improved were those who initially missed the CM principle but were able to invoke it after browsing over the solved problem. Students in the algebra-based course had greater difficulty invoking the CM principle if they were not provided with the solved problem, and the improvement in the performance of the intervention groups as compared to that of the comparison group in the algebra-based course was mainly due to this fact. Being able to *invoke* all the relevant principles, however, is not enough. In order to solve the problem correctly, students must be able to *apply* the principles correctly, which requires an ability to decompose a multi-step problem into several sub-problems and understanding how different sub-problems are connected. Our previous research [1, 2] indicates that students are able to perform analogical reasoning between two problems both of which have two steps and can be solved by applying the same principles in the same order. Adding an additional step to the problem increases the difficulty significantly and transfer becomes challenging. We hypothesize that by giving students more explicit guidance and practice on how to divide and connect the sub-problems and how to learn and organize the information from the solved problem, students will gradually develop expertise. They will learn about the applicability of physics principles in diverse situations and the coherence of the knowledge structure in physics if such analogical reasoning activities are sustained and rewarded throughout the course.

## REFERENCES

1. Lin and Singh, submitted 2010 PERC Proceedings.
2. E. Yerushalmi, A. Mason, E. Cohen, and C. Singh, AIP Conf. Proc. **1179**, 23-26, (2009).
3. Singh and Rosengrant, Am. J. Phys. **71**, 607-617 (2003).